\documentclass[letter]{aa}  

\usepackage{graphicx}
\usepackage{subcaption}
\usepackage{txfonts}
\usepackage{hyperref}
\usepackage{xcolor}
\usepackage{tikz,xcolor}
\usepackage{csquotes}
\usepackage{siunitx}

\usepackage{longtable}
\usepackage[para]{threeparttable}
\usepackage[para]{threeparttablex}
\usepackage{placeins}
\usepackage{multicol}

\usepackage{scalerel}
\usepackage{tikz}
\usetikzlibrary{svg.path}

\definecolor{orcidlogocol}{HTML}{A6CE39}
\tikzset{
  orcidlogo/.pic={
    \fill[orcidlogocol] svg{M256,128c0,70.7-57.3,128-128,128C57.3,256,0,198.7,0,128C0,57.3,57.3,0,128,0C198.7,0,256,57.3,256,128z};
    \fill[white] svg{M86.3,186.2H70.9V79.1h15.4v48.4V186.2z}
                 svg{M108.9,79.1h41.6c39.6,0,57,28.3,57,53.6c0,27.5-21.5,53.6-56.8,53.6h-41.8V79.1z M124.3,172.4h24.5c34.9,0,42.9-26.5,42.9-39.7c0-21.5-13.7-39.7-43.7-39.7h-23.7V172.4z}electronically
                 svg{M88.7,56.8c0,5.5-4.5,10.1-10.1,10.1c-5.6,0-10.1-4.6-10.1-10.1c0-5.6,4.5-10.1,10.1-10.1C84.2,46.7,88.7,51.3,88.7,56.8z};
  }
}

\newcommand\orcidicon[1]{\href{https://orcid.org/#1}{\mbox{\scalerel*{
\begin{tikzpicture}[yscale=-1,transform shape]
\pic{orcidlogo};
\end{tikzpicture}
}{|}}}}

\newcommand{\review}[1]{#1} %Paper Review
\begin{document}

   \title{Constraints on the X-ray-to-radio fluence ratio of FRB\,20240114A}

   \subtitle{}
   \author{F.\,Eppel\inst{1,2}%\orcidicon{0000-0001-7112-9942}
          \and
          M.\,Krumpe\inst{3}
          \and
          P.\,Limaye\inst{4,2}
          \and
          N.\,Intrarat\inst{5}
          \and
          J.\,Wongphechauxsorn\inst{1,2}
          \and
          M.\,Cruces\inst{6,7,8,9,2}
          \and
          W.\,Herrmann\inst{10}
          \and
          F.\,Jankowski\inst{11}
          \and
          P.\,Jaroenjittichai\inst{5}
          \and
          L. G.\,Spitler\inst{2}
          \and
          M.\,Kadler\inst{1}
          }

   \institute{Julius-Maximilians-Universität Würzburg, Institut für Theoretische Physik und Astrophysik, Lehrstuhl für Astronomie, Emil-Fischer-Straße 31, D-97074 Würzburg, Germany\\
              \email{florian.eppel@uni-wuerzburg.de}
         \and
         Max-Planck-Institut für Radioastronomie, Auf dem Hügel 69, D-53121 Bonn, Germany
         \and
        Leibniz-Institut für Astrophysik Potsdam (AIP), An der Sternwarte 16, D-14482 Potsdam, Germany
        \and
        Argelander Institute for Astronomy, Auf Dem Hügel 71, D-53121 Bonn, Germany 
        \and
        National Astronomical Research Institute of Thailand, Chiang Mai, Thailand
        \and
        Centre of Astro-Engineering, Pontificia Universidad Catolica de Chile, Av. Vicuna Mackenna 4860, Santiago, Chile
        \and
        Department of Electrical Engineering, Pontificia Universidad Catolica de Chile, Av. Vicuna Mackenna 4860, Santiago, Chile
        \and
        European Southern Observatory, Karl-Schwarzschild-Str. 2, D-85748 Garching bei München, Germany
        \and
        Joint ALMA Observatory, Alonso de Córdova 3107, Vitacura, Santiago, Chile
        \and
        Astropeiler Stockert e.V., Astropeiler 2-4, D-53902 Bad Münstereifel, Germany
        \and
        LPC2E, OSUC, Univ Orleans, CNRS, CNES, Observatoire de Paris, F-45071 Orleans, France
        }

   \date{Received Dec 20, 2024; accepted Feb 5, 2025}

% \abstract{}{}{}{}{} 
% 5 {} token are mandatory
 
  \abstract
  % context heading (optional)
  {We report on multiwavelength observations of FRB\,20240114A, a nearby (z=0.13), hyperactive, repeating fast radio burst that was discovered in January 2024. We performed simultaneous observations of the source with the Effelsberg 100-m radio telescope, the Thai National Radio Telescope, the Astropeiler Stockert, and the X-ray satellite \textsl{XMM-Newton} in May 2024. On May 23, 2024, we detected 459 bursts from the source using the Ultra-Broad-Band (UBB) receiver of the Effelsberg telescope, covering a frequency range from 1.3\,GHz to 6\,GHz. All bursts have simultaneous X-ray coverage, which allows us to put stringent constraints on the X-ray-to-radio fluence ratio, $\eta_{x/r}$, of FRB\,20240114A. In this work, we focus on
   the three brightest radio bursts detected during the campaign. The brightest burst exhibits a radio fluence of $1.4\times 10^{-17}$\,erg\,cm$^{-2}$, while the $3\sigma$ upper limit of the 0.2$-$12\,keV  absorption-corrected X-ray burst fluence lies in the range of $3.4\times 10^{-11}$\,erg\,cm$^{-2}$ to $1.7\times 10^{-10}$\,erg\,cm$^{-2}$, depending on the spectral model. Assuming a  10\,keV black-body spectrum, the X-ray-to-radio fluence ratio can be constrained to $\eta_{x/r}<1.2\times10^{7}$. 
   A cutoff power law  ($\Gamma=1.56$, cutoff at 84\,keV) is also considered, physically motivated by the Galactic magnetar SGR\,1935+2154, which has previously shown X-ray bursts associated with FRB-like radio bursts at a measured X-ray-to-radio fluence ratio of $\eta_{x/r}\sim2.5\times 10^{5}$ (1$-$250\,keV). In this scenario, we find that $\eta_{x/r}<2.4\times 10^6$. Our results are consistent with FRB\,20240114A being powered by a  mechanism similar to that of SGR\,1935+2154. We show that future multiwavelength campaigns will be able to improve this limit if sufficiently bright radio bursts are observed with simultaneous X-ray coverage.} 

   \keywords{X-rays: bursts - Stars: magnetars - Radiation mechanisms: nonthermal - Methods: observational - Methods: data analysis}

\authorrunning{F.\,Eppel et al.}

\maketitle
%
%-------------------------------------------------------------------

\nolinenumbers
\section{Introduction}

The origin of fast radio bursts (FRBs) is still unclear and a plethora of theoretical models have been suggested to explain these bright pulses of radio emission with millisecond duration \citep{platts2018}. Several of the proposed models, especially those involving magnetars, predict associated multiwavelength \review{(MWL)} emission \citep[e.g.,][]{Metzger2019,Lu}. However, previous searches for X-ray or $\gamma$-ray counterparts of FRBs did not lead to any detection \citep{Scholz2016,Scholz2017,Scholz2020,Pilia2020,Piro2021,Trudu2023,Pearlman2023,Yan2024,Cook2024}. In early 2020, X-ray telescopes succeeded in detecting a high-energy transient signal coincident with an FRB-like burst \citep{Tavani2021,Mereghetti,Bochenek2020,Li2021}. The source of this signal was identified as the Galactic magnetar SGR\,1935+2154, which exhibited several more simultaneous radio/X-ray flares \citep[e.g.,][]{Ridnaia}. This discovery showed for the first time that a magnetar can produce X-ray bursts in coincidence with FRB-like radio bursts, as had been predicted before \citep[e.g.,][]{Metzger2019}. It further suggests that coordinated radio and X-ray observations of FRBs may be able to detect coincident signals from other bright magnetar flares also in nearby extragalactic systems.

\noindent 
On January 26, 2024, the CHIME/FRB collaboration reported the discovery of the repeating FRB\,20240114A \citep{AtelChimeFRB20240114A}. Multiple follow-up observations throughout the following weeks found very bright bursts \citep{ATel16432,ATel16547,ATel16494,ATel16452,ATel16433}, burst storms of several hundred bursts \citep{ATel16505,ATel16430,ATel16565}, and high-frequency radio detections up to 6\,GHz \citep{ATel16599,ATel16597,ATel16620}. %Moreover, there were reports about coincident gamma-ray emission from the position of FRB\,20240114A \citep{ATel16594,ATel16630}. However, the significance of this detection was disputed by \cite{ATel16602}. 
FRB\,20240114A was localized to the dwarf star-forming galaxy SDSS\,J212739.84+041945.8 with sub-arcsecond precision \citep{MeerKATLocalization,EVNFRB20240114A} at a redshift of $z=0.130\pm 0.002$ \citep{ATel16613}.

Motivated by the high-fluence bursts, hyperactivity, and the possibility of detecting or constraining associated X-ray emission, we obtained target-of-opportunity observations of FRB\,20240114A with the \textsl{XMM-Newton} X-ray satellite using director's discretionary time (DDT). Additionally, we obtained simultaneous DDT radio observations with the Effelsberg 100-m telescope, the Thai National Radio Telescope (TNRT), and the Stockert Astropeiler. Here, we present the first results from this \review{MWL} campaign. In Sect.\,\ref{sec:obs}, we give an overview of the observations and analysis for all telescopes, in Sect.\,\ref{sec:results} we present results for the three brightest bursts detected in the campaign, and in Sect.\,\ref{sec:discussion} we discuss our findings in the broader context of FRB \review{MWL} studies.

\begin{figure}
    \centering
    \includegraphics[width=\linewidth]{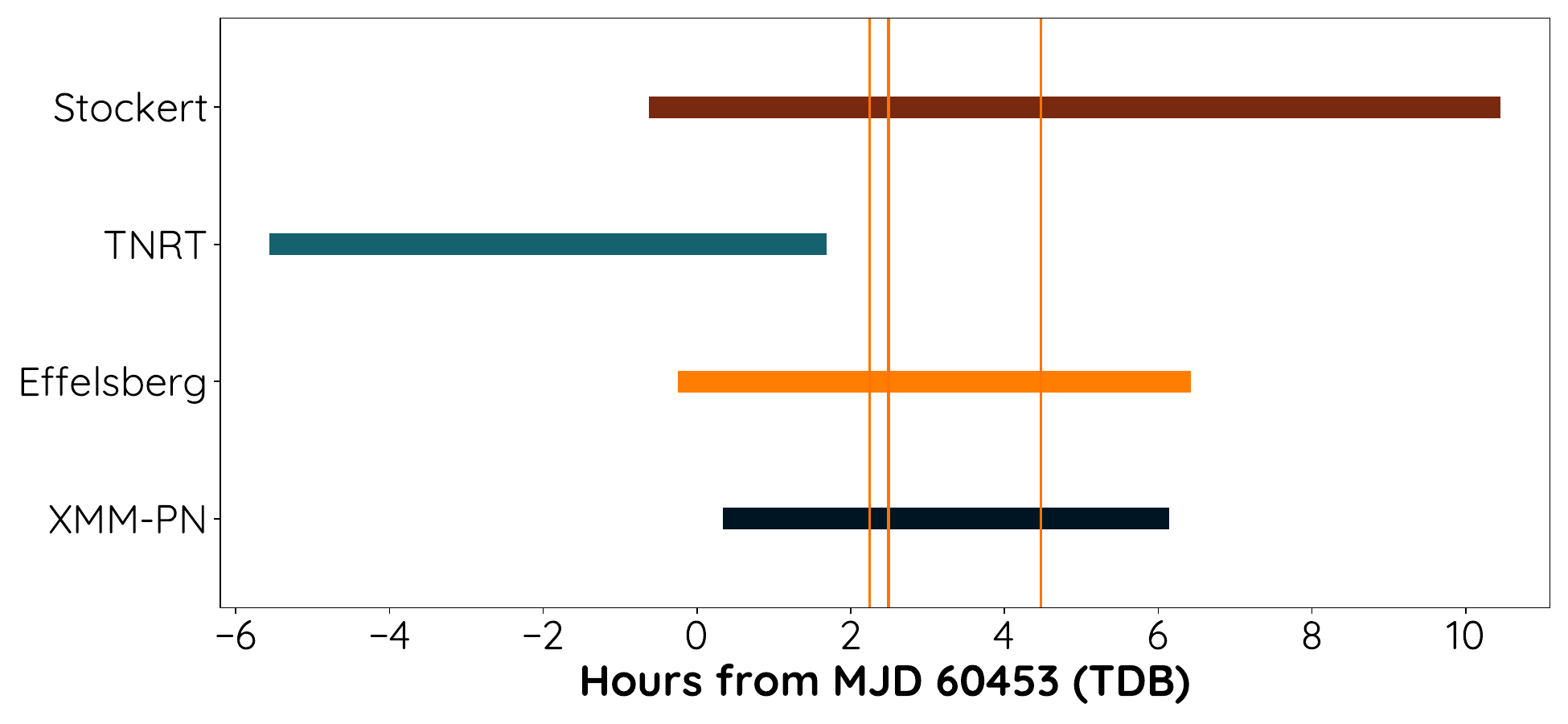}
    \caption{Overview of the observing times (barycentric) and telescopes on May 23, 2024. For \textsl{XMM-Newton}-PN, the displayed time range corresponds to the GTIs. The three vertical lines indicate the arrival times of burst B1, B2, and B3.}
    \label{fig:obsdetail}
\end{figure}

%--------------------------------------------------------------------
\section{Observations and analysis}
\label{sec:obs}

The observations were split into three sessions. Two sessions were carried out with simultaneous observations by Stockert and \textsl{XMM-Newton} (May 17, 2024, and May 27, 2024) and one session (May 23, 2024) included TNRT, Effelsberg, Stockert, and \textsl{XMM-Newton}. An overview of the observing times on May 23, 2024, is shown in Fig.\,\ref{fig:obsdetail}. In the following, we provide an overview of the observing setups and analysis performed for all telescopes.

\subsection{Effelsberg 100-m Telescope}
We observed FRB\,20240114A with the Effelsberg 100-m telescope on May 23, 2024, for $\sim$7 hours using the Ultra-Broad-Band (UBB) receiver, which covers a frequency range of 1.3$-$6\,GHz \review{with a system equivalent flux density (SEFD) of 12$-$16\,Jy}\footnote{\url{https://eff100mwiki.mpifr-bonn.mpg.de/doku.php?id=information_for_astronomers:rx:p170mm}}. The data were recorded using the Effelsberg Direct Digitisation (EDD) backend \citep{EDD} with a channel width of 0.5\,MHz and time resolution of 128\,$\mu$s. The data were recorded in five sub-bands (band 1: 1.3$-$1.9\,GHz, band 2: 1.9$-$2.6\,GHz, band 3: 3.0$-$4.1\,GHz, band 4: 4.1$-$5.2\,GHz, and band 5: 5.2$-$6.0\,GHz). We searched the data for single bursts using \texttt{TransientX} \citep{TransientX} separately in every sub-band and found 459 unique bursts that could be identified as FRBs matched across sub-bands. For a detailed overview of the UBB data analysis, we refer to Appendix \ref{appendix:ubb}.

\subsection{Stockert Telescope}

The Astropeiler Stockert is a 25-m telescope operating in the L band. The SEFD of the instrument is 385 Jy. Data was recorded as total-intensity 32-bit data with 98 MHz bandwidth (1332.25\,MHz to 1430.5\,MHz) and a time resolution of 218.45\,$\mu$s using a fast Fourier transform backend \citep{pffts}. 
%Data was converted from the instrument specific format to filterbank files for further analysis. 
We searched for radio bursts using \texttt{prepsubband} and \texttt{single\_pulse\_search} from \texttt{PRESTO} \citep{PRESTO}. The threshold for single pulses was set to a \review{signal-to-noise ratio (S/N)} of 7, corresponding to a fluence threshold of $\sim 14$\,Jy\,ms for 1\,ms wide pulses. The default boxcar size with 30 time bins \review{(i.e., $\sim 6.5$\,ms)} was used for the search. The data were filtered for RFI from nearby radar systems by applying the modulation index technique described by \cite{Spitler_2012} and then classified using \texttt{FETCH} \citep{Fetch1}. The bursts flagged by \texttt{FETCH} as potential candidates were manually inspected. During the time of the observation reported here, no burst was detected. An additional manual inspection of the data was performed around the arrival time of \review{the brightest Effelsberg} burst (B1) with no convincing result. It has to be noted that on other observing days, bright burst (i.e., $> {12}$\,Jy\,ms) were detected using the analysis described above.
 
\subsection{Thai National Radio Telescope}

The TNRT is a 40-m radio telescope located at Doi Saket, Chiang Mai, Thailand, equipped with an EDD backend\footnote{\url{https://indico.narit.or.th/event/197/page/654-tnro-40-m-tnrt}}.
%(18°51'52" N latitude and 99°13'01" E longitude), 400m above sea level\footnotemark{} \footnotetext{see }. 
We performed pulsar search mode observations of FRB\,20240114A on 23 May 2024 for $\sim 7$ hours (see Fig.\,\ref{fig:obsdetail}), using the L-band receiver centered at 1350 MHz with a bandwidth of 900 MHz, and time resolution of 51.2\,$\mu$s. 
%Due to hardware limitations,  no coherent dedispersion has been done. 
%The observation started on 23 May 2024 at 1:43 AM and finished at 8:30 AM local time (GMT+7), see Figure \ref{fig:obsdetail}.
%We used \texttt{TransientX} to search for single-pulses with DM between 510 and 570 $\textrm{cm}^{-3} \textrm{pc}$ and \texttt{replot\_fil} to filter out bad candidates with similar settings as for the Effelsberg analysis. No significant burst was detected.
We used \texttt{TransientX} to search for single-pulses with DM 510$-$570\,$\textrm{cm}^{-3}\,\textrm{pc}$. The S/N cutoff was set to 7 and pulse width threshold $<30$\,ms. After inspecting all candidates visually, no burst \review{was} found. This is consistent with the expected burst rate of 0.1 to 0.3 bursts/hour, assuming a conservative SEFD of 90\,Jy \citep{TNRTwhitepaper}.

\subsection{XMM-Newton}

We obtained \textsl{XMM-Newton} DDT observations that were carried out in three different time slots, on May 17, 2024 (Obs.ID 0935190601, 25.2\,ks total observing time), May 23, 2024 (Obs.ID 0935190701, 25\,ks), and May 27, 2024 (\review{Obs.ID} 0935191001, 28.7\,ks). We used the PN- and MOS-instruments with medium filter in small window mode to achieve the best timing accuracy while also having good spatial resolution.

The data for the PN- and MOS-instruments were processed using the \textsl{XMM-Newton} Science Analysis System \citep[\texttt{SAS},][]{XMMSAS} version 21.0.0. We used the pipelines \texttt{epproc} for the PN-camera and \texttt{emproc} for the MOS-cameras to generate filtered event lists. We performed the barycentric correction using the \texttt{barycen} task. Additionally, we identified times of flaring background by analyzing the light curves at 10$-$15\,keV energies. By applying conservative cuts to the background count rate, we identified the final good time intervals (GTIs) to use for the analysis of the persistent flux limit, totaling 57.3\,ks. The event lists were extracted in the energy band 0.2$-$12\,keV. Additionally, we performed an astrometric pointing correction using the \texttt{eposcorr} task (see Appendix \ref{appendix:eposcorr}).

\begin{figure*}
    \centering
    \includegraphics[width=0.33\linewidth]{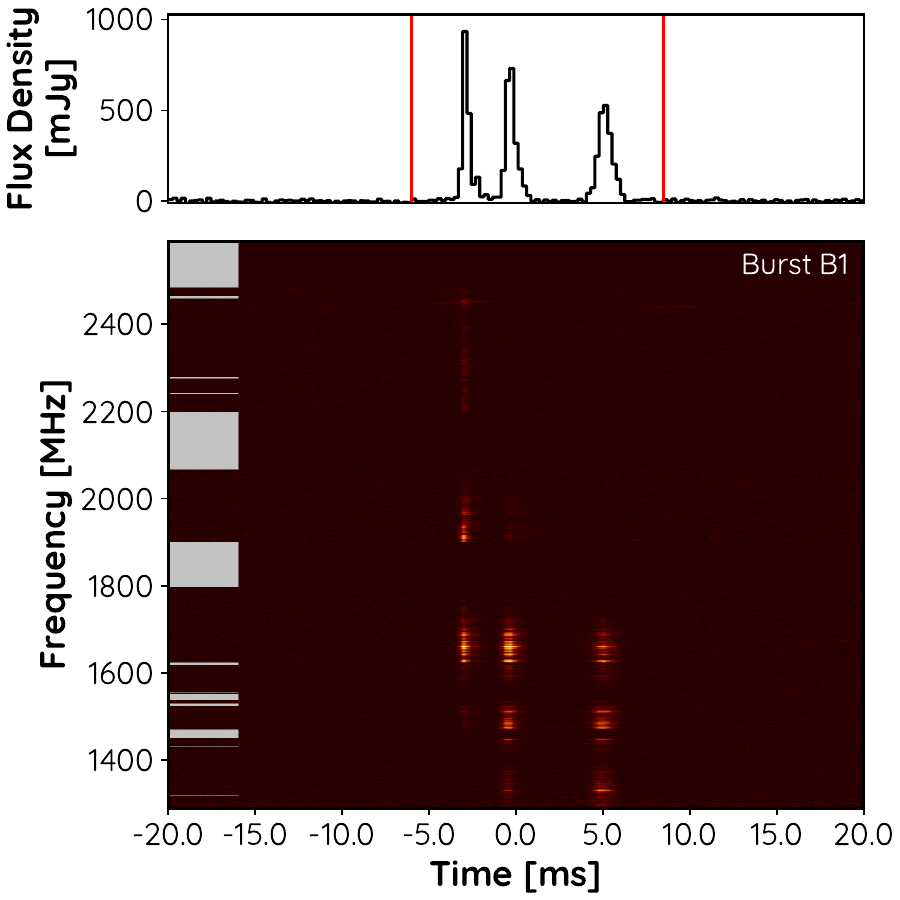}
    \includegraphics[width=0.33\linewidth]{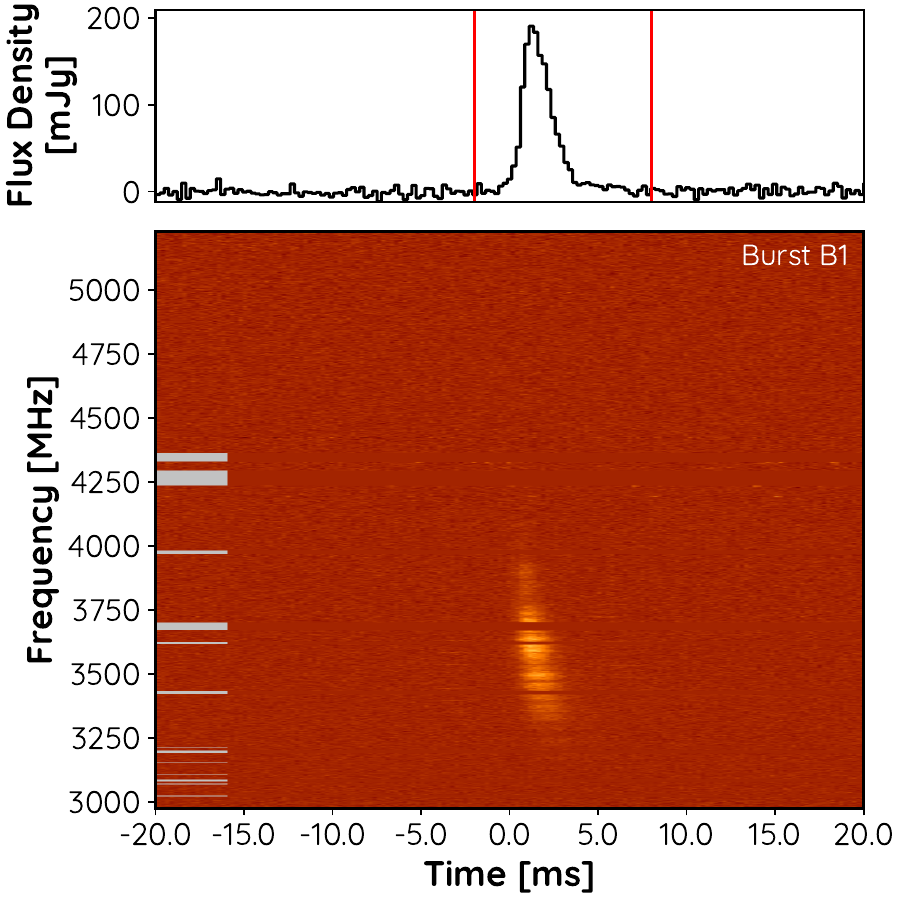}
    \includegraphics[width=0.33\linewidth]{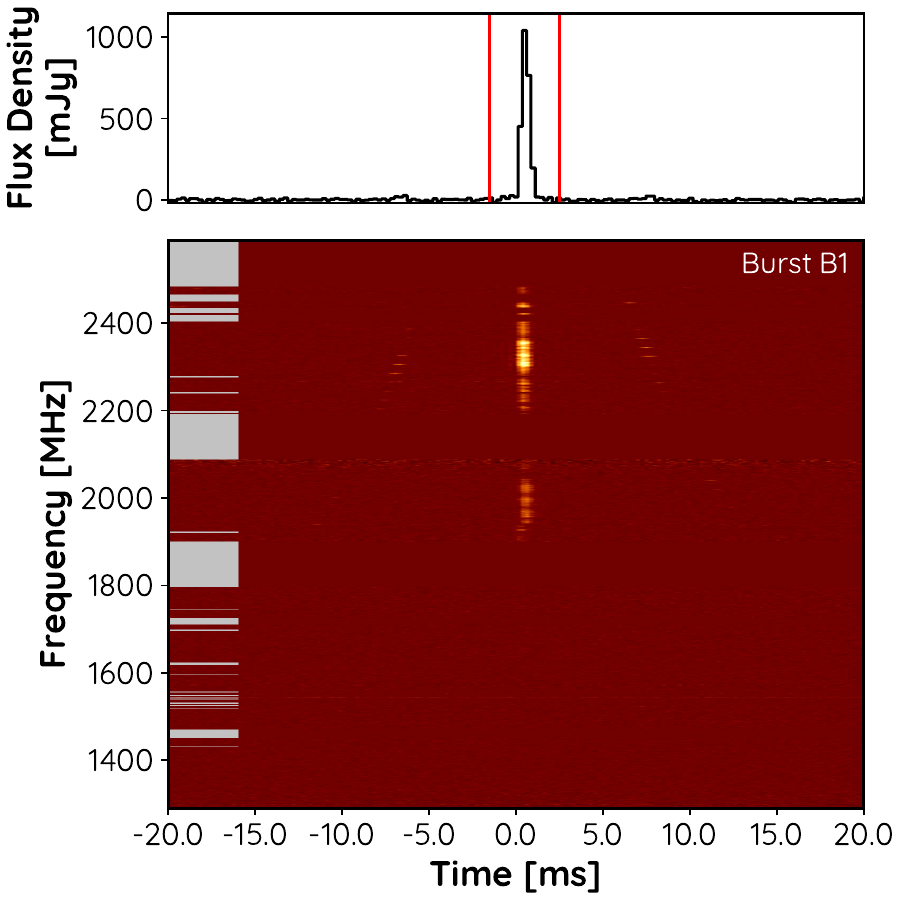}    \caption{Waterfall plots of the three brightest bursts from FRB\,20240114A that were detected simultaneous to the \textsl{XMM-Newton} observation on May 23, 2024 (left: B1, center: B2, right: B3). RFI-flagged channels are highlighted by the gray lines on the left. The top panel shows the flux density, averaged over the displayed bandwidth, excluding RFI-flagged channels, \review{and the time windows used for the fluence calculation (red lines)}. The displayed bandwidth for B1 and B3 \review{covers} UBB sub-bands 1 and 2, while B2 is shown in UBB sub-bands 3 and 4. The color maps are scaled individually for every burst from minimum to maximum flux density.}
    \label{fig:bursts}
\end{figure*}

\section{Results}
\label{sec:results}

\subsection{Radio bursts}
During the Effelsberg observation on May 23, 2024, we detected 459 bursts from FRB\,20240114A. None of these bursts were seen by TNRT or Stockert. \review{In UBB band 1, which overlaps with TNRT and Stockert, we detected 186 bursts, all except B1 with S/N\,$<$\,50. The brightest burst overlapping with TNRT was detected with S/N\,=\,34.2. Given the sensitivity of Effelsberg, which is superior by a factor of $\sim6$ to TNRT and $\sim25$ to Stockert, and its broader frequency coverage, the non-detection with Stockert and TNRT is expected.} In the additional two \textsl{XMM-Newton} sessions that were covered only by Stockert, no bursts were found.

In this work, we present the three brightest bursts (labeled B1, B2, and B3) detected at an S/N~$>$100 with Effelsberg, which are shown in Fig.\,\ref{fig:bursts} as waterfall plots. \review{In the case of a constant X-ray-to-radio fluence ratio, $\eta_{x/r}$, these are the best candidates to constrain possibly associated X-ray emission.} We combined all sub-bands of the UBB receiver in which significant emission of the burst was detected. For burst B1 and B3, this corresponds to sub-band 1 and 2, while B2 shows significant emission in sub-bands 3 and 4. Burst B1 shows temporal substructure, divided into three sub-bursts that follow the characteristic so-called sad-trombone scheme \citep[e.g.,][]{sadtrombone}. Bursts B2 and B3 are single bursts without visible subcomponents. To calculate the fluence of each burst, we first averaged the flux density over the observing bandwidth, excluding RFI-flagged channels and then integrated the flux density over a \review{time window covering the full burst (see Fig.\,\ref{fig:bursts})}. The resulting fluences are shown in Table\,\ref{tab:fluences}. In order to compare the radio burst properties with X-ray observations, we calculated the bandwidth-integrated fluence, also referred to as the \enquote{radio burst fluence}, $\mathcal{F_\mathrm{R}}$ \citep[e.g.][]{Pearlman2023,Cook2024}. This quantity is calculated by integrating the fluence over the observing bandwidth (excluding RFI-flagged channels).
%This quantity is much more useful to compare the energy release of the burst with multi-wavelength observations, since it is independent of the bandwidth that was used to calculate the fluence, if the burst bandwidth is fully captured. 

By using the known redshift of the host galaxy of FRB\,20240114A \citep[$z=0.13$,][]{ATel16613} and assuming a flat Lambda-CDM model \citep{Planck2020}, we calculated the intrinsic radio burst release energy, $\mathrm{E}_\mathrm{R}$, for every burst, following \cite{Zhang2018}. The  values obtained are displayed in Table\,\ref{tab:fluences}. The brightest burst in our observations, B1, summed over all three subcomponents, exhibits a radio burst fluence of $\mathcal{F_\mathrm{R,B1}}=1.4\times 10^{-17}\,\mathrm{erg}\,\mathrm{cm^{-2}}$, which translates to a total release energy of $\mathrm{E}_\mathrm{R,B1}=\review{5.8}\times 10^{38}\,\textrm{erg}$. We note that all of these values must be understood as lower limits, since parts of the burst emission exceed the receiver bandwidth (at least for B1) and multiple channels within the burst bandwidth needed to be flagged due to RFI and could therefore not be considered.

\begin{table*}
\centering
\begin{threeparttable}
\caption{\small Overview of the three brightest bursts detected within our MWL campaign.}
\centering
\begin{tabular}{c@{\,}ccc@{\,}ccc@{\,}c}
\hline\hline
{Burst-ID}  & {TOA\tnote{a}} & {Fluence\tnote{b}} & {Bandwidth\tnote{c}} & {Radio Burst\tnote{d}} & {Radio Burst\tnote{e}} & {X-ray Burst\tnote{f}} & {Fluence Ratio\tnote{g}} \\
  {}   &  {[MJD]}  & {[Jy ms]} & {[MHz]}  & {Fluence [erg cm$^{-2}$]}  & {Energy [erg]} & {Fluence [erg cm$^{-2}$]} & {$\eta_\mathrm{x/r}$} \\
\hline
B1 & 60453.093476719 & 1.\review{55}  & 881 & $1.4\times 10^{-17}$ & $\review{5.8}\times10^{38}$ & $<3.4\times 10^{-11}$ & $<2.4\times 10^{6}$ \\
B2 & 60453.103822424 & 0.3\review{8} & 2015 & $7.\review{6}\times 10^{-18}$ & $3.2\times10^{38}$ & $<3.4\times 10^{-11}$ & $<4.5\times 10^{6}$ \\
B3 & 60453.186311513 & 0.\review{63} & 830 & $5.\review{2}\times 10^{-18}$ & $2.\review{2}\times10^{38}$ & $<3.4\times 10^{-11}$ & $<\review{6.5}\times 10^{6}$ \\
\hline
\end{tabular}
\label{tab:fluences}
\begin{tablenotes}
    \tnote{a}~Time of arrival corrected to infinite frequency at the solar system barycenter.
    \tnote{b}~Fluence averaged over the observing bandwidth.
    \tnote{c}~Effective observing bandwidth used to calculate the average fluence, excluding flagged channels.
    \tnote{d}~Radio burst fluence, $\mathcal{F_\mathrm{R}}$ (i.e., Fluence $\times$ Bandwidth).
    \tnote{e}~Redshift-corrected intrinsic total radio burst release energy.
    \tnote{f}~X-ray burst fluence, $\mathcal{F}_\mathrm{X}$ (0.2$-$12\,keV, corrected for Galactic absorption), 3\,$\sigma$ upper limit, assuming a cutoff power law spectrum as was observed for SGR\,1935+2154 \citep{Li2021}.
    \tnote{g}~3$\sigma$ upper limit to the X-ray (0.2$-$12\,keV) to radio fluence ratio ($\eta_\mathrm{x/r}=\mathcal{F}_\mathrm{X}/\mathcal{F}_\mathrm{R}$).
\end{tablenotes}
\end{threeparttable}
\end{table*}

\subsection{X-ray burst emission}
\label{sec:burstsxray}

We checked the PN-event list (0.2$-$12\,keV) for X-ray photons at the times of radio bursts B1$-$B3. We used an extraction region with a radius of 0.01$^\circ$ (i.e., 90\,\% encircled energy fraction at 1.5\,keV) centered on the position of FRB\,20240114A \citep[RA = 21h27m39.835s, Dec = +04d19m45.634s;][]{EVNFRB20240114A}. \review{This analysis was done on the non-background-corrected data with an} average background count rate in this region of $\sim 3\times\,10^{-2}$\,cps. For burst B1, the closest X-ray photon after the burst was detected 48\,s after the radio emission, for burst B2 50\,s after the radio emission, and for burst B3 17\,s. This is consistent with the expected background rate; hence, we conclude that none of the bursts are detected significantly at X-ray energies. 

In order to derive an upper limit on the burst X-ray flux for each of the radio bursts, we used a similar method as described by \cite{Cook2024}. We estimated the background rate at the position of FRB\,20240114A using a 200\,s window centered on the time of arrival of the burst. By applying low-count photon statistics using the \texttt{pwkit} library \citep{pwkit}, following \cite{Kraft}, we derived a 3\,$\sigma$ (i.e., 99.73\,\% one-sided confidence level) upper limit of 5.9\,photons at the time of the radio burst, or 6.5\,photons when correcting for the \review{90\,\%} encircled energy fraction. Considering a typical duration of magnetar-associated X-ray bursts of $\sim$100\,ms as seen in SGR\,1935+2154 \citep{Mereghetti}, a cutoff power law spectrum ($\Gamma=1.56$, cutoff 84\,keV) as observed in SGR\,1935+2154 \citep{Li2021}, and an average Galactic hydrogen column density of $N_\textrm{H}=5.0\times10^{20}\,\mathrm{cm}^{-2}$ in the direction of FRB\,20240114A \citep{columndensity}, this translates to a 3\,$\sigma$ upper limit to the X-ray fluence of $\mathcal{F}_\mathrm{X}<3.4\times10^{-11}\,\mathrm{erg}\,\mathrm{cm}^{-2}$ in the 0.2$-$12\,keV band, corrected for Galactic absorption \citep[following Appendix A of][]{Tubin}. The corresponding upper limit on the intrinsic total X-ray energy release is $\mathrm{E}_{X}<1.5\times10^{45}$\,erg (0.2$-$12\,keV). Assuming a 10\,keV black-body spectrum \citep[cf.,][]{Cook2024}, the 3\,$\sigma$ upper limit to the X-ray fluence is $\mathcal{F}_\mathrm{X}\,<\,1.7\times\,10^{-10}\,\mathrm{erg}\,\mathrm{cm}^{-2}$ ($\mathrm{E}_\mathrm{X}<7.2\times10^{45}$\,erg) in the 0.2$-$12\,keV band, corrected for Galactic absorption. 

We used the radio burst fluences, $\mathcal{F}_\mathrm{R}$, and the upper limit on the X-ray burst fluence, $\mathcal{F}_\mathrm{X}$, to calculate upper limits on the X-ray-to-radio fluence ratio, $\eta_\mathrm{x/r}=\mathcal{F}_\mathrm{X}/\mathcal{F}_\mathrm{R}$.
%which is a useful quantity to constrain existing FRB models and to compare with multi-wavelength observations of other FRB sources. 
This fluence ratio always corresponds to specific observing bandwidths in the radio and X-ray. The fluence ratios for the three brightest bursts in our observing campaign are shown in Table\,\ref{tab:fluences}. The most constraining limit can be obtained from burst B1, yielding a fluence ratio of $\eta_\mathrm{x/r}<2.4\times 10^6$ at 3\,$\sigma$ confidence, assuming a spectrum similar to SGR\,1935+2154. The \review{assumption of a} 10\,keV black-body spectrum leads to \review{a less stringent limit of} $\eta_{x/r}<1.2\times 10^7$.

\subsection{Persistent X-ray emission}

To constrain the X-ray flux of a possible persistent source associated with FRB\,20240114A, we selected events coming from its position with a radius of 0.01$^\circ$ in an energy range of 0.2$-$12\,keV. We defined background regions next to the position of the FRB without any X-ray sources to compare the number of collected photons at the position of the FRB with the background expectation. \review{Following} \cite{Kraft}, we calculated a $3\,\sigma$ upper limit on the persistent X-ray flux \review{using} 57.3\,ks of GTI PN-data. We find a 3\,$\sigma$ upper limit of $<2.4\times 10^{-3}$ cps, corrected for the encircled energy fraction. Assuming a power law spectrum with $\Gamma=2$ and the same $N_\textrm{H}$ as in Sect.\,\ref{sec:burstsxray}, this corresponds to a persistent X-ray flux limit of $<1.0\times 10^{-14}\,\mathrm{erg}\,\mathrm{cm}^{-2}\,\mathrm{s}^{-1}$ and an isotropic luminosity of $\mathcal{L}_\mathrm{X}<4.3\times 10^{41}\,\mathrm{erg}\,\mathrm{s^{-1}}$ (0.2$-$12\,keV). 

%Itimproves the upper limits reported by \cite{Atel16645} based on \textsl{Swift} observations by a factor of 10$-$100.

\section{Discussion}
\label{sec:discussion}

\subsection{Comparison with other multiwavelength FRB studies}
In this section, we compare our results to \review{MWL} observations of other FRB\review{s}, mainly focusing on a recent publication by \cite{Cook2024}, which includes an overview of many previous studies. We note that in all previous studies, no significant X-ray emission (either persistent or burst-like) was found. \cite{Cook2024} used X-ray fluences and energies in the 0.5$-$10\,keV band. We therefore converted our limits on the X-ray burst fluence from Sect.\,\ref{sec:results} to \review{this} band. In \review{the 0.5$-$10\,keV} energy range, the upper limit 
%to the persistent X-ray flux of FRB\,20240114A corresponds to $6.9\times 10^{-15}\,\mathrm{erg}\,\mathrm{cm}^{-2}\,\mathrm{s}^{-1}$, while the upper limit 
on the X-ray burst fluence corresponds to $2.7\times 10^{-11}\,\mathrm{erg}\,\mathrm{cm}^{-2}$ (cutoff power law) or $1.1\times 10^{-10}\,\mathrm{erg}\,\mathrm{cm}^{-2}$ (black-body), assuming the same spectral models as in Sect.\,\ref{sec:results}. 

%While \cite{Cook2024} observed a different source slightly closer to Earth (FRB\,20220912A, $z=0.077$), the \textsl{XMM-Newton} background photon count rates are comparable to the values that we find in our analysis. For the persistent X-ray flux, \cite{Cook2024} derive an upper limit of $8.8\times 10^{-15}\,\mathrm{erg}\,\mathrm{cm}^{-2}\,\mathrm{s}^{-1}$, which is slightly larger than the limit we find. This discrepancy can be explained by the different hydrogen column densities assumed in the calculation. While \cite{Cook2024} use a conservative value of $N_\textrm{H}=10^{22}\,\mathrm{cm}^{-2}$, trying to account for absorption at the source, we use the average hydrogen column density of $N_\textrm{H}=5.0\times10^{20}\,\mathrm{cm}^{-2}$ in the direction of FRB\,20240114A \citep{columndensity}. If we assume the same hydrogen density as \cite{Cook2024}, our upper limit to the persistent X-ray flux of FRB\,20240114A increases to $1.6\times 10^{-14}\,\mathrm{erg}\,\mathrm{cm}^{-2}\,\mathrm{s}^{-1}$ in the 0.5\,keV$-$10\,keV band. This value is higher than the limit found by \cite{Cook2024}, which can be explained by the different background count rates in our observation compared to the \cite{Cook2024} observation.

For the single burst X-ray fluence limits at the time of the radio bursts, we used a similar analysis as presented by \cite{Cook2024}. Their upper limits are $\sim\,35\,\%$ higher for the black-body spectrum scenario, which can be explained by the different $N_\textrm{H}$ values assumed. 
%We note that without accounting for internal source absorption, the hydrogen column density in the direction of FRB\,20220912A is by a factor of two higher than for FRB\,20240114A. Therefore, the slightly lower upper limits in our analysis are naturally expected since less X-ray absorption is present. 
While \cite{Cook2024} use a conservative value of $N_\textrm{H}=10^{22}\,\mathrm{cm}^{-2}$, trying to account for absorption at the source, we use the average hydrogen column density of $N_\textrm{H}=5.0\times10^{20}\,\mathrm{cm}^{-2}$ in the direction of FRB\,20240114A \citep{columndensity}.
Assuming the more conservative value of $N_\textrm{H}=10^{22}\,\mathrm{cm}^{-2}$, our upper limit for the single burst X-ray fluence increases to $1.2\times 10^{-10}\,\mathrm{erg}\,\mathrm{cm}^{-2}$ for a black-body spectral model (0.5$-$10\,keV). This value is still slightly lower \review{than} the limit obtained by \cite{Cook2024}, which is likely due to the fact that we initially extracted a larger energy band of 0.2$-$12\,keV from the \textsl{XMM-Newton} data and then downsampled to the 0.5$-$10\,keV band. Moreover, it is unclear if \cite{Cook2024} used the same count-to-flux conversion as we did, based on \cite{Tubin}.

According to \cite{Cook2024}, the most constraining fluence ratio limit (0.5$-$10\,keV) for a single FRB to date was found at $\eta_\mathrm{x/r}<7\times 10^6$ for FRB\,20220912A and FRB\,20180916B \citep{Pilia2020}, assuming a 10\,keV black-body X-ray spectrum. Our best limit obtained from the brightest burst B1 ($\eta_\mathrm{x/r}<2.4\times 10^6$) is three times lower than this, but considers a different X-ray energy band (0.2$-$12\,keV), different spectral model (cutoff power law), and different absorption. 

\review{While the assumption of a blackbody spectrum is motivated by observations of hard X-ray bursts from magnetars \citep[e.g., ][]{blackbody1,blackbody2}, X-ray emission from SGR\,1935+2154 suggests a nonthermal cutoff power law \citep{Li2021}. It is thus unclear which spectrum should be assumed.} \review{Using a blackbody model like \cite{Cook2024},} the limit from burst B1 transforms to $\eta_\mathrm{x/r}<8.5\times 10^6$ at 0.5$-$10\,keV. This is comparable to the lowest single burst limits found for FRB\,20220912A and FRB\,20180916B.
The limit can be further constrained by applying a stacking approach \citep[e.g.,][]{Cook2024}; however, this requires an accurate flux calibration of all 459 bursts and will be done in a separate publication.

\subsection{Constraints on \review{magnetars as progenitors}}
With an X-ray-to-radio fluence ratio of $\eta_{x/r}<2.4 \times 10^{6}$, we cannot rule out the currently favored FRB production models, based on magnetars. As pointed out by \cite{Cook2024}, most magnetar-based models predict $\eta_\mathrm{x/r}\approx 10^0-10^4$ \citep[e.g.,][]{Margalit,Lu,Popov}. Moreover, the X-ray-to-radio fluence ratio from the Galactic magnetar SGR\,1935+2154 was observed to be $\eta_\mathrm{x/r}\sim2.5\times 10^{5}$ \citep[1$-$250\,keV,][]{Bochenek2020,Li2021}, which is still well below the most constraining upper limit for $\eta_\mathrm{x/r}$ found for any extragalactic FRB. Corrected for typical \textsl{XMM-Newton} observing bands (0.2$-$12\,keV, \review{cutoff power law, $\Gamma=1.56$, cutoff 84\,keV}), the expected $\eta_{x/r}$ from SGR\,1935+2154 is even lower, by a factor of $\sim3$. Therefore, our observations are consistent with FRB\,20240114A being powered by a magnetar, in agreement with synchrotron maser models \citep[e.g.][]{Metzger2019,Margalit} and magnetospheric emission models \citep[e.g.,][]{Lu, Zhong2024}. \review{However, it is unlikely that FRB\,20240114A exhibits pulsed radio emission with anti-aligned X-ray pulses as seen in SGR\,1935+2154 \citep{zhu}, which requires $\eta_{x/r}\sim(7-12)\times 10^6$ \citep{Cook2024}.}

\section{Summary and outlook} 
\label{sec:conclusions}

We observed the repeating FRB\,20240114A with the Effelsberg 100-m telescope, the \review{TNRT}, the Astropeiler Stockert, and the X-ray satellite \textsl{XMM-Newton}. On May 23, 2024, we detected 459 radio bursts from FRB\,20240114A using the Effelsberg UBB receiver, covering a frequency range from 1.3$-$6\,GHz. In this work, we have presented the three brightest bursts, detected with S/N~$>100$. The brightest burst (B1, see Fig.\,\ref{fig:bursts}) exhibits a radio fluence of $1.4\times 10^{-17}\,\mathrm{erg}\,\mathrm{cm}^{-2}$. None of the bursts were detected with \textsl{XMM-Newton}, Stockert, or TNRT. We used the \textsl{XMM-Newton} observations to derive upper fluence limits for associated X-ray bursts lasting 100\,ms, which can be constrained to $\mathcal{F}_\mathrm{X}<3.4\times10^{-11}\,\mathrm{erg}\,\mathrm{cm}^{-2}$ in the 0.2$-$12\,keV band at 3\,$\sigma$ confidence. This corresponds to an X-ray-to-radio fluence ratio of $\eta_{x/r}<2.4\times 10^{6}$ for the brightest burst. Considering different magnetar models for FRBs, which predict associated X-ray emission \citep[e.g.][]{Metzger2019,Margalit,Lu,Zhong2024}, this limit is consistent with FRB\,20240114A being powered by a magnetar. Moreover, from the \review{MWL} detection of radio and X-ray bursts from the Galactic magnetar SGR\,1935+2154, one expects $\eta_\mathrm{x/r}\sim2.5\times 10^{5}$ \citep[1-250\,keV,][]{Bochenek2020,Li2021}, which is in agreement with our limit. That means FRB\,20240114A could be powered by a similar mechanism as SGR\,1935+2154.
In a follow-up publication, we are planning to stack all 459 bursts to derive an even lower limit for the X-ray/radio fluence ratio. Given the extremely bright bursts from FRB\,20240114A reported, for example, by \cite{AtelChimeFRB20240114A}, and its hyperactivity, it is a prime target for future \review{MWL} studies. While future studies with current X-ray telescopes will not be able to place significantly lower X-ray flux limits, the detection of more energetic radio bursts up to a factor of 10$-$100 times brighter than the ones found in our campaign is possible \citep[e.g.,][]{Kirsten2024} and would result in superior constraints on $\eta_\mathrm{x/r}$, reaching the value found for SGR\,1935+2154. 
%Albeit logistically much more difficult to set up, systematic multi-wavelength searches for the possibly more energetic once-off FRBs \citep{Chen2022} should also be considered, especially since it is currently unclear if they are of the same nature as repeating FRBs \citep[cf.][]{Kirsten2024}.

\begin{acknowledgements}
\review{We thank the anonymous referee for valuable comments that helped us improve the manuscript.}
We thank Amanda Cook for helpful discussions on the X-ray count-to-flux conversion. This work is based on observations with the 100-m telescope of the MPIfR (Max-Planck-Institut für Radioastronomie) at Effelsberg. FE and MKa acknowledge support from the Deutsche Forschungsgemeinschaft (DFG, grant 447572188), and MKr from the German Aerospace Center (DLR, grant FKZ 50 OR 2307). LGS is a Lise Meitner research group leader and acknowledges support from the Max Planck Society. MC is a Max-Planck partner group leader and acknowledges support from Max-Planck society. Based on observations obtained with XMM-Newton, an ESA science mission with instruments and contributions directly funded by ESA Member States and NASA. This research is partially based on observations made with the 40-m Thai National Radio Telescope at the Thai National Radio Astronomy Observatory under the open-use program ID ToO1 (Target of Opportunity 1), which is operated by the National Astronomical Research Institute of Thailand (Public Organization). We thank the NARIT and MPIfR teams for providing the new capabilities used at TNRT and Effelsberg. The TNRT L-band receiver and USB system were developed
in collaboration between the MPIfR and NARIT. The Effelsberg UBB receiver and
EDD system are developed and maintained by the electronics division at the
MPIfR. %This research has made use of the Astrophysics Data System, funded by NASA under Cooperative Agreement 80NSSC21M00561.
\end{acknowledgements}

\bibliographystyle{aa} % style aa.bst
\bibliography{bib} % your references Yourfile.bib

\begin{thebibliography}{63}
\expandafter\ifx\csname natexlab\endcsname\relax\def\natexlab#1{#1}\fi

\bibitem[{{Agarwal} {et~al.}(2020){Agarwal}, {Aggarwal}, {Burke-Spolaor}, {Lorimer}, \& {Garver-Daniels}}]{Fetch1}
{Agarwal}, D., {Aggarwal}, K., {Burke-Spolaor}, S., {Lorimer}, D.~R., \& {Garver-Daniels}, N. 2020, \mnras, 497, 1661

\bibitem[{{An} {et~al.}(2014){An}, {Kaspi}, {Beloborodov}, {Kouveliotou}, {Archibald}, {Boggs}, {Christensen}, {Craig}, {Gotthelf}, {Grefenstette}, {Hailey}, {Harrison}, {Madsen}, {Mori}, {Stern}, \& {Zhang}}]{blackbody1}
{An}, H., {Kaspi}, V.~M., {Beloborodov}, A.~M., {et~al.} 2014, \apj, 790, 60

\bibitem[{{Barr} {et~al.}(2023){Barr}, {Bansod}, {Behrend}, {Esser}, {Kasemann}, {Wieching}, {Winchen}, \& {Wu}}]{EDD}
{Barr}, E.~D., {Bansod}, A., {Behrend}, J., {et~al.} 2023, in 2023 XXXVth General Assembly and Scientific Symposium of the International Union of Radio Science (URSI GASS), We--J01--PM3--4

\bibitem[{{Barr} {et~al.}(2013){Barr}, {Champion}, {Kramer}, {Eatough}, {Freire}, {Karuppusamy}, {Lee}, {Verbiest}, {Bassa}, {Lyne}, {Stappers}, {Lorimer}, \& {Klein}}]{pffts}
{Barr}, E.~D., {Champion}, D.~J., {Kramer}, M., {et~al.} 2013, \mnras, 435, 2234

\bibitem[{{Bhardwaj} {et~al.}(2024){Bhardwaj}, {Kirichenko}, \& {Gil de Paz}}]{ATel16613}
{Bhardwaj}, M., {Kirichenko}, A., \& {Gil de Paz}, A. 2024, The Astronomer's Telegram, 16613

\bibitem[{{Bochenek} {et~al.}(2020){Bochenek}, {Ravi}, {Belov}, {Hallinan}, {Kocz}, {Kulkarni}, \& {McKenna}}]{Bochenek2020}
{Bochenek}, C.~D., {Ravi}, V., {Belov}, K.~V., {et~al.} 2020, \nat, 587, 59

\bibitem[{{Cook} {et~al.}(2024){Cook}, {Scholz}, {Pearlman}, {Abbott}, {Cruces}, {Gaensler}, {Dong}, {Michilli}, {Eadie}, {Kaspi}, {Stairs}, {Tan}, {Bhardwaj}, {Cassanelli}, {Curtin}, {Ibik}, {Lazda}, {Masui}, {Pandhi}, {Rafiei-Ravandi}, {Sammons}, {Shin}, {Smith}, \& {Stenning}}]{Cook2024}
{Cook}, A.~M., {Scholz}, P., {Pearlman}, A.~B., {et~al.} 2024, \apj, 974, 170

\bibitem[{{Coti Zelati} {et~al.}(2021){Coti Zelati}, {Borghese}, {Israel}, {Rea}, {Esposito}, {Pilia}, {Burgay}, {Possenti}, {Corongiu}, {Ridolfi}, {Dehman}, {Vigan{\`o}}, {Turolla}, {Zane}, {Tiengo}, \& {Keane}}]{blackbody2}
{Coti Zelati}, F., {Borghese}, A., {Israel}, G.~L., {et~al.} 2021, \apjl, 907, L34

\bibitem[{{Gabriel} {et~al.}(2004){Gabriel}, {Denby}, {Fyfe}, {Hoar}, {Ibarra}, {Ojero}, {Osborne}, {Saxton}, {Lammers}, \& {Vacanti}}]{XMMSAS}
{Gabriel}, C., {Denby}, M., {Fyfe}, D.~J., {et~al.} 2004, in Astronomical Society of the Pacific Conference Series, Vol. 314, Astronomical Data Analysis Software and Systems (ADASS) XIII, ed. F.~{Ochsenbein}, M.~G. {Allen}, \& D.~{Egret}, 759

\bibitem[{{Gaia Collaboration} {et~al.}(2023){Gaia Collaboration}, {Vallenari}, {Brown}, {Prusti}, {de Bruijne}, {Arenou}, {Babusiaux}, {Biermann}, {Creevey}, {Ducourant}, {Evans}, {Eyer}, {Guerra}, {Hutton}, {Jordi}, {Klioner}, {Lammers}, {Lindegren}, {Luri}, {Mignard}, {Panem}, {Pourbaix}, {Randich}, {Sartoretti}, {Soubiran}, {Tanga}, {Walton}, {Bailer-Jones}, {Bastian}, {Drimmel}, {Jansen}, {Katz}, {Lattanzi}, {van Leeuwen}, {Bakker}, {Cacciari}, {Casta{\~n}eda}, {De Angeli}, {Fabricius}, {Fouesneau}, {Fr{\'e}mat}, {Galluccio}, {Guerrier}, {Heiter}, {Masana}, {Messineo}, {Mowlavi}, {Nicolas}, {Nienartowicz}, {Pailler}, {Panuzzo}, {Riclet}, {Roux}, {Seabroke}, {Sordo}, {Th{\'e}venin}, {Gracia-Abril}, {Portell}, {Teyssier}, {Altmann}, {Andrae}, {Audard}, {Bellas-Velidis}, {Benson}, {Berthier}, {Blomme}, {Burgess}, {Busonero}, {Busso}, {C{\'a}novas}, {Carry}, {Cellino}, {Cheek}, {Clementini}, {Damerdji}, {Davidson}, {de Teodoro}, {Nu{\~n}ez Campos}, {Delchambre}, {Dell'Oro}, {Esquej},
  {Fern{\'a}ndez-Hern{\'a}ndez}, {Fraile}, {Garabato}, {Garc{\'\i}a-Lario}, {Gosset}, {Haigron}, {Halbwachs}, {Hambly}, {Harrison}, {Hern{\'a}ndez}, {Hestroffer}, {Hodgkin}, {Holl}, {Jan{\ss}en}, {Jevardat de Fombelle}, {Jordan}, {Krone-Martins}, {Lanzafame}, {L{\"o}ffler}, {Marchal}, {Marrese}, {Moitinho}, {Muinonen}, {Osborne}, {Pancino}, {Pauwels}, {Recio-Blanco}, {Reyl{\'e}}, {Riello}, {Rimoldini}, {Roegiers}, {Rybizki}, {Sarro}, {Siopis}, {Smith}, {Sozzetti}, {Utrilla}, {van Leeuwen}, {Abbas}, {{\'A}brah{\'a}m}, {Abreu Aramburu}, {Aerts}, {Aguado}, {Ajaj}, {Aldea-Montero}, {Altavilla}, {{\'A}lvarez}, {Alves}, {Anders}, {Anderson}, {Anglada Varela}, {Antoja}, {Baines}, {Baker}, {Balaguer-N{\'u}{\~n}ez}, {Balbinot}, {Balog}, {Barache}, {Barbato}, {Barros}, {Barstow}, {Bartolom{\'e}}, {Bassilana}, {Bauchet}, {Becciani}, {Bellazzini}, {Berihuete}, {Bernet}, {Bertone}, {Bianchi}, {Binnenfeld}, {Blanco-Cuaresma}, {Blazere}, {Boch}, {Bombrun}, {Bossini}, {Bouquillon}, {Bragaglia}, {Bramante}, {Breedt},
  {Bressan}, {Brouillet}, {Brugaletta}, {Bucciarelli}, {Burlacu}, {Butkevich}, {Buzzi}, {Caffau}, {Cancelliere}, {Cantat-Gaudin}, {Carballo}, {Carlucci}, {Carnerero}, {Carrasco}, {Casamiquela}, {Castellani}, {Castro-Ginard}, {Chaoul}, {Charlot}, {Chemin}, {Chiaramida}, {Chiavassa}, {Chornay}, {Comoretto}, {Contursi}, {Cooper}, {Cornez}, {Cowell}, {Crifo}, {Cropper}, {Crosta}, {Crowley}, {Dafonte}, {Dapergolas}, {David}, {David}, {de Laverny}, {De Luise}, {De March}, {De Ridder}, {de Souza}, {de Torres}, {del Peloso}, {del Pozo}, {Delbo}, {Delgado}, {Delisle}, {Demouchy}, {Dharmawardena}, {Di Matteo}, {Diakite}, {Diener}, {Distefano}, {Dolding}, {Edvardsson}, {Enke}, {Fabre}, {Fabrizio}, {Faigler}, {Fedorets}, {Fernique}, {Fienga}, {Figueras}, {Fournier}, {Fouron}, {Fragkoudi}, {Gai}, {Garcia-Gutierrez}, {Garcia-Reinaldos}, {Garc{\'\i}a-Torres}, {Garofalo}, {Gavel}, {Gavras}, {Gerlach}, {Geyer}, {Giacobbe}, {Gilmore}, {Girona}, {Giuffrida}, {Gomel}, {Gomez}, {Gonz{\'a}lez-N{\'u}{\~n}ez},
  {Gonz{\'a}lez-Santamar{\'\i}a}, {Gonz{\'a}lez-Vidal}, {Granvik}, {Guillout}, {Guiraud}, {Guti{\'e}rrez-S{\'a}nchez}, {Guy}, {Hatzidimitriou}, {Hauser}, {Haywood}, {Helmer}, {Helmi}, {Sarmiento}, {Hidalgo}, {Hilger}, {H{\l}adczuk}, {Hobbs}, {Holland}, {Huckle}, {Jardine}, {Jasniewicz}, {Jean-Antoine Piccolo}, {Jim{\'e}nez-Arranz}, {Jorissen}, {Juaristi Campillo}, {Julbe}, {Karbevska}, {Kervella}, {Khanna}, {Kontizas}, {Kordopatis}, {Korn}, {K{\'o}sp{\'a}l}, {Kostrzewa-Rutkowska}, {Kruszy{\'n}ska}, {Kun}, {Laizeau}, {Lambert}, {Lanza}, {Lasne}, {Le Campion}, {Lebreton}, {Lebzelter}, {Leccia}, {Leclerc}, {Lecoeur-Taibi}, {Liao}, {Licata}, {Lindstr{\o}m}, {Lister}, {Livanou}, {Lobel}, {Lorca}, {Loup}, {Madrero Pardo}, {Magdaleno Romeo}, {Managau}, {Mann}, {Manteiga}, {Marchant}, {Marconi}, {Marcos}, {Marcos Santos}, {Mar{\'\i}n Pina}, {Marinoni}, {Marocco}, {Marshall}, {Martin Polo}, {Mart{\'\i}n-Fleitas}, {Marton}, {Mary}, {Masip}, {Massari}, {Mastrobuono-Battisti}, {Mazeh}, {McMillan}, {Messina}, {Michalik},
  {Millar}, {Mints}, {Molina}, {Molinaro}, {Moln{\'a}r}, {Monari}, {Mongui{\'o}}, {Montegriffo}, {Montero}, {Mor}, {Mora}, {Morbidelli}, {Morel}, {Morris}, {Muraveva}, {Murphy}, {Musella}, {Nagy}, {Noval}, {Oca{\~n}a}, {Ogden}, {Ordenovic}, {Osinde}, {Pagani}, {Pagano}, {Palaversa}, {Palicio}, {Pallas-Quintela}, {Panahi}, {Payne-Wardenaar}, {Pe{\~n}alosa Esteller}, {Penttil{\"a}}, {Pichon}, {Piersimoni}, {Pineau}, {Plachy}, {Plum}, {Poggio}, {Pr{\v{s}}a}, {Pulone}, {Racero}, {Ragaini}, {Rainer}, {Raiteri}, {Rambaux}, {Ramos}, {Ramos-Lerate}, {Re Fiorentin}, {Regibo}, {Richards}, {Rios Diaz}, {Ripepi}, {Riva}, {Rix}, {Rixon}, {Robichon}, {Robin}, {Robin}, {Roelens}, {Rogues}, {Rohrbasser}, {Romero-G{\'o}mez}, {Rowell}, {Royer}, {Ruz Mieres}, {Rybicki}, {Sadowski}, {S{\'a}ez N{\'u}{\~n}ez}, {Sagrist{\`a} Sell{\'e}s}, {Sahlmann}, {Salguero}, {Samaras}, {Sanchez Gimenez}, {Sanna}, {Santove{\~n}a}, {Sarasso}, {Schultheis}, {Sciacca}, {Segol}, {Segovia}, {S{\'e}gransan}, {Semeux}, {Shahaf}, {Siddiqui}, {Siebert},
  {Siltala}, {Silvelo}, {Slezak}, {Slezak}, {Smart}, {Snaith}, {Solano}, {Solitro}, {Souami}, {Souchay}, {Spagna}, {Spina}, {Spoto}, {Steele}, {Steidelm{\"u}ller}, {Stephenson}, {S{\"u}veges}, {Surdej}, {Szabados}, {Szegedi-Elek}, {Taris}, {Taylor}, {Teixeira}, {Tolomei}, {Tonello}, {Torra}, {Torra}, {Torralba Elipe}, {Trabucchi}, {Tsounis}, {Turon}, {Ulla}, {Unger}, {Vaillant}, {van Dillen}, {van Reeven}, {Vanel}, {Vecchiato}, {Viala}, {Vicente}, {Voutsinas}, {Weiler}, {Wevers}, {Wyrzykowski}, {Yoldas}, {Yvard}, {Zhao}, {Zorec}, {Zucker}, \& {Zwitter}}]{GaiaDR3}
{Gaia Collaboration}, {Vallenari}, A., {Brown}, A.~G.~A., {et~al.} 2023, \aap, 674, A1

\bibitem[{{Hessels} {et~al.}(2019){Hessels}, {Spitler}, {Seymour}, {Cordes}, {Michilli}, {Lynch}, {Gourdji}, {Archibald}, {Bassa}, {Bower}, {Chatterjee}, {Connor}, {Crawford}, {Deneva}, {Gajjar}, {Kaspi}, {Keimpema}, {Law}, {Marcote}, {McLaughlin}, {Paragi}, {Petroff}, {Ransom}, {Scholz}, {Stappers}, \& {Tendulkar}}]{FRB121102Hessels}
{Hessels}, J.~W.~T., {Spitler}, L.~G., {Seymour}, A.~D., {et~al.} 2019, \apjl, 876, L23

\bibitem[{{Hewitt} {et~al.}(2024){Hewitt}, {Huang}, {Hessels}, {Cognard}, {Guillemot}, {Ould-Boukattine}, {Snelders}, \& {Kirsten}}]{ATel16597}
{Hewitt}, D.~M., {Huang}, J., {Hessels}, J.~W.~T., {et~al.} 2024, The Astronomer's Telegram, 16597

\bibitem[{{HI4PI Collaboration} {et~al.}(2016){HI4PI Collaboration}, {Ben Bekhti}, {Fl{\"o}er}, {Keller}, {Kerp}, {Lenz}, {Winkel}, {Bailin}, {Calabretta}, {Dedes}, {Ford}, {Gibson}, {Haud}, {Janowiecki}, {Kalberla}, {Lockman}, {McClure-Griffiths}, {Murphy}, {Nakanishi}, {Pisano}, \& {Staveley-Smith}}]{columndensity}
{HI4PI Collaboration}, {Ben Bekhti}, N., {Fl{\"o}er}, L., {et~al.} 2016, \aap, 594, A116

\bibitem[{{Jaroenjittichai} {et~al.}(2022){Jaroenjittichai}, {Sugiyama}, {Kramer}, {Soonthornthum}, {Akahori}, {Asanok}, {Baan}, {Bran}, {Breen}, {Cho}, {Chanapote}, {Dodson}, {Ellingsen}, {Etoka}, {Gray}, {Green}, {Hada}, {Halson}, {Hirota}, {Honma}, {Imai}, {Johnston}, {Kim}, {Kramer}, {Li}, {Macatangay}, {Menten}, {Minh}, {Mkrtichian}, {Pimpanuwat}, {Richards}, {Rioja}, {Rujopakarn}, {Sakai}, {Sakai}, {Samanso}, {Sanpa-arsa}, {Semenko}, {Sunada}, {Surapipith}, {Thoonsaengngam}, {Voronkov}, {Wongphecauxson}, {Kesh Yadav}, {Zhang}, {Zheng}, \& {Poshyachinda}}]{TNRTwhitepaper}
{Jaroenjittichai}, P., {Sugiyama}, K., {Kramer}, B.~H., {et~al.} 2022, arXiv e-prints, arXiv:2210.04926

\bibitem[{{Joshi} {et~al.}(2024){Joshi}, {Medina}, {Earwicker}, {Farah}, {Gajjar}, {Sheikh}, {Pollak}, {Siemion}, {Cruz}, {Hickish}, {Premnath}, {DeBoer}, {Donnachie}, {Singh}, {Davis}, {Snodgrass}, \& {Karn}}]{ATel16599}
{Joshi}, P., {Medina}, A., {Earwicker}, J.~T., {et~al.} 2024, The Astronomer's Telegram, 16599

\bibitem[{{Katz}(2023)}]{sadtrombone}
{Katz}, J.~I. 2023, \mnras, 518, 2015

\bibitem[{{Kirsten} {et~al.}(2024){Kirsten}, {Ould-Boukattine}, {Herrmann}, {Gawro{\'n}ski}, {Hessels}, {Lu}, {Snelders}, {Chawla}, {Yang}, {Blaauw}, {Nimmo}, {Puchalska}, {Wolak}, \& {van Ruiten}}]{Kirsten2024}
{Kirsten}, F., {Ould-Boukattine}, O.~S., {Herrmann}, W., {et~al.} 2024, Nature Astronomy, 8, 337

\bibitem[{{Kraft} {et~al.}(1991){Kraft}, {Burrows}, \& {Nousek}}]{Kraft}
{Kraft}, R.~P., {Burrows}, D.~N., \& {Nousek}, J.~A. 1991, \apj, 374, 344

\bibitem[{{Kumar} {et~al.}(2024){Kumar}, {Maan}, \& {Bhusare}}]{ATel16452}
{Kumar}, A., {Maan}, Y., \& {Bhusare}, Y. 2024, The Astronomer's Telegram, 16452

\bibitem[{{Kurpas} {et~al.}(2024){Kurpas}, {Schwope}, {Pires}, \& {Haberl}}]{kurpas2024}
{Kurpas}, J., {Schwope}, A.~D., {Pires}, A.~M., \& {Haberl}, F. 2024, \aap, 683, A164

\bibitem[{{Li} {et~al.}(2021){Li}, {Lin}, {Xiong}, {Ge}, {Li}, {Li}, {Lu}, {Zhang}, {Tuo}, {Nang}, {Zhang}, {Xiao}, {Chen}, {Song}, {Xu}, {Liu}, {Jia}, {Cao}, {Qu}, {Zhang}, {Gu}, {Liao}, {Zhao}, {Tan}, {Nie}, {Zhao}, {Zheng}, {Zheng}, {Luo}, {Cai}, {Li}, {Xue}, {Bu}, {Chang}, {Chen}, {Chen}, {Chen}, {Chen}, {Chen}, {Cui}, {Cui}, {Deng}, {Dong}, {Du}, {Fu}, {Gao}, {Gao}, {Gao}, {Gu}, {Guan}, {Guo}, {Han}, {Huang}, {Huo}, {Jiang}, {Jiang}, {Jin}, {Jin}, {Kong}, {Li}, {Li}, {Li}, {Li}, {Li}, {Li}, {Li}, {Liang}, {Liu}, {Liu}, {Liu}, {Liu}, {Liu}, {Lu}, {Lu}, {Luo}, {Ma}, {Meng}, {Ou}, {Sai}, {Shang}, {Song}, {Sun}, {Tao}, {Wang}, {Wang}, {Wang}, {Wang}, {Wang}, {Wen}, {Wu}, {Wu}, {Wu}, {Xiao}, {Xu}, {Yang}, {Yang}, {Yang}, {Yang}, {Yi}, {Yin}, {You}, {Zhang}, {Zhang}, {Zhang}, {Zhang}, {Zhang}, {Zhang}, {Zhang}, {Zhang}, {Zhang}, {Zhang}, {Zhang}, {Zhang}, {Zhang}, {Zhang}, {Zhang}, {Zhang}, {Zhou}, {Zhou}, {Zhu}, {Zhu}, \& {Zhuang}}]{Li2021}
{Li}, C.~K., {Lin}, L., {Xiong}, S.~L., {et~al.} 2021, Nature Astronomy, 5, 378

\bibitem[{{Limaye} \& {Spitler}(2024)}]{ATel16620}
{Limaye}, P. \& {Spitler}, L. 2024, The Astronomer's Telegram, 16620

\bibitem[{{Lu} {et~al.}(2020){Lu}, {Kumar}, \& {Zhang}}]{Lu}
{Lu}, W., {Kumar}, P., \& {Zhang}, B. 2020, \mnras, 498, 1397

\bibitem[{{Luo} {et~al.}(2021){Luo}, {Ransom}, {Demorest}, {Ray}, {Archibald}, {Kerr}, {Jennings}, {Bachetti}, {van Haasteren}, {Champagne}, {Colen}, {Phillips}, {Zimmerman}, {Stovall}, {Lam}, \& {Jenet}}]{PINT}
{Luo}, J., {Ransom}, S., {Demorest}, P., {et~al.} 2021, \apj, 911, 45

\bibitem[{{Margalit} {et~al.}(2020){Margalit}, {Beniamini}, {Sridhar}, \& {Metzger}}]{Margalit}
{Margalit}, B., {Beniamini}, P., {Sridhar}, N., \& {Metzger}, B.~D. 2020, \apjl, 899, L27

\bibitem[{{Men} \& {Barr}(2024)}]{TransientX}
{Men}, Y. \& {Barr}, E. 2024, \aap, 683, A183

\bibitem[{{Mereghetti} {et~al.}(2020){Mereghetti}, {Savchenko}, {Ferrigno}, {G{\"o}tz}, {Rigoselli}, {Tiengo}, {Bazzano}, {Bozzo}, {Coleiro}, {Courvoisier}, {Doyle}, {Goldwurm}, {Hanlon}, {Jourdain}, {von Kienlin}, {Lutovinov}, {Martin-Carrillo}, {Molkov}, {Natalucci}, {Onori}, {Panessa}, {Rodi}, {Rodriguez}, {S{\'a}nchez-Fern{\'a}ndez}, {Sunyaev}, \& {Ubertini}}]{Mereghetti}
{Mereghetti}, S., {Savchenko}, V., {Ferrigno}, C., {et~al.} 2020, \apjl, 898, L29

\bibitem[{{Metzger} {et~al.}(2019){Metzger}, {Margalit}, \& {Sironi}}]{Metzger2019}
{Metzger}, B.~D., {Margalit}, B., \& {Sironi}, L. 2019, \mnras, 485, 4091

\bibitem[{{Morello} {et~al.}(2019){Morello}, {Barr}, {Cooper}, {Bailes}, {Bates}, {Bhat}, {Burgay}, {Burke-Spolaor}, {Cameron}, {Champion}, {Eatough}, {Flynn}, {Jameson}, {Johnston}, {Keith}, {Keane}, {Kramer}, {Levin}, {Ng}, {Petroff}, {Possenti}, {Stappers}, {van Straten}, \& {Tiburzi}}]{clfd}
{Morello}, V., {Barr}, E.~D., {Cooper}, S., {et~al.} 2019, \mnras, 483, 3673

\bibitem[{{Ould-Boukattine} {et~al.}(2024{\natexlab{a}}){Ould-Boukattine}, {Dijkema}, {Gawronski}, {Herrmann}, {Hessels}, {Kirsten}, {Snelders}, {Beer}, {Bijlsma}, {Blaauw}, {Boons}, {Boven}, {Buchsteiner}, {Engelskirchen}, {Fischer}, {Hewitt}, {Loge}, {Marcote}, {van der Meer}, {Mulder}, {Munk}, {Nitsche}, {Ovinge}, {Puchalska}, {Sanders}, {Schmitz}, {Sluman}, {Telkamp}, {Wolf}, \& {Yang}}]{ATel16565}
{Ould-Boukattine}, O.~S., {Dijkema}, T.~J., {Gawronski}, M., {et~al.} 2024{\natexlab{a}}, The Astronomer's Telegram, 16565

\bibitem[{{Ould-Boukattine} {et~al.}(2024{\natexlab{b}}){Ould-Boukattine}, {Hessels}, {Kirsten}, {Hewitt}, {Snelders}, {Blaauw}, {Sluman}, {Mulder}, {Herrmann}, {Gawronski}, {Puchalska}, \& {Gopinath}}]{ATel16432}
{Ould-Boukattine}, O.~S., {Hessels}, J.~W.~T., {Kirsten}, F., {et~al.} 2024{\natexlab{b}}, The Astronomer's Telegram, 16432

\bibitem[{{Panda} {et~al.}(2024){Panda}, {Bhattacharyya}, {Dudeja}, {Kudale}, \& {Roy}}]{ATel16494}
{Panda}, U., {Bhattacharyya}, S., {Dudeja}, C., {Kudale}, S., \& {Roy}, J. 2024, The Astronomer's Telegram, 16494

\bibitem[{{Pearlman} {et~al.}(2025){Pearlman}, {Scholz}, {Bethapudi}, {Hessels}, {Kaspi}, {Kirsten}, {Nimmo}, {Spitler}, {Fonseca}, {Meyers}, {Stairs}, {Tan}, {Bhardwaj}, {Chatterjee}, {Cook}, {Curtin}, {Dong}, {Eftekhari}, {Gaensler}, {G{\"u}ver}, {Kaczmarek}, {Leung}, {Masui}, {Michilli}, {Prince}, {Sand}, {Shin}, {Smith}, \& {Tendulkar}}]{Pearlman2023}
{Pearlman}, A.~B., {Scholz}, P., {Bethapudi}, S., {et~al.} 2025, Nature Astronomy, 9, 111

\bibitem[{{Pelliciari} {et~al.}(2024){Pelliciari}, {Geminardi}, {Bernardi}, {Pilia}, {Esposito}, \& {Naldi}}]{ATel16547}
{Pelliciari}, D., {Geminardi}, A., {Bernardi}, G., {et~al.} 2024, The Astronomer's Telegram, 16547

\bibitem[{{Perley} \& {Butler}(2017)}]{perleybutler2017}
{Perley}, R.~A. \& {Butler}, B.~J. 2017, \apjs, 230, 7

\bibitem[{{Pilia} {et~al.}(2020){Pilia}, {Burgay}, {Possenti}, {Ridolfi}, {Gajjar}, {Corongiu}, {Perrodin}, {Bernardi}, {Naldi}, {Pupillo}, {Ambrosino}, {Bianchi}, {Burtovoi}, {Casella}, {Casentini}, {Cecconi}, {Ferrigno}, {Fiori}, {Gendreau}, {Ghedina}, {Naletto}, {Nicastro}, {Ochner}, {Palazzi}, {Panessa}, {Papitto}, {Pittori}, {Rea}, {Castillo}, {Savchenko}, {Setti}, {Tavani}, {Trois}, {Trudu}, {Turatto}, {Ursi}, {Verrecchia}, \& {Zampieri}}]{Pilia2020}
{Pilia}, M., {Burgay}, M., {Possenti}, A., {et~al.} 2020, \apjl, 896, L40

\bibitem[{{Piro} {et~al.}(2021){Piro}, {Bruni}, {Troja}, {O'Connor}, {Panessa}, {Ricci}, {Zhang}, {Burgay}, {Dichiara}, {Lee}, {Lotti}, {Niu}, {Pilia}, {Possenti}, {Trudu}, {Xu}, {Zhu}, {Kutyrev}, \& {Veilleux}}]{Piro2021}
{Piro}, L., {Bruni}, G., {Troja}, E., {et~al.} 2021, \aap, 656, L15

\bibitem[{{Planck Collaboration} {et~al.}(2020){Planck Collaboration}, {Aghanim}, {Akrami}, {Ashdown}, {Aumont}, {Baccigalupi}, {Ballardini}, {Banday}, {Barreiro}, {Bartolo}, {Basak}, {Battye}, {Benabed}, {Bernard}, {Bersanelli}, {Bielewicz}, {Bock}, {Bond}, {Borrill}, {Bouchet}, {Boulanger}, {Bucher}, {Burigana}, {Butler}, {Calabrese}, {Cardoso}, {Carron}, {Challinor}, {Chiang}, {Chluba}, {Colombo}, {Combet}, {Contreras}, {Crill}, {Cuttaia}, {de Bernardis}, {de Zotti}, {Delabrouille}, {Delouis}, {Di Valentino}, {Diego}, {Dor{\'e}}, {Douspis}, {Ducout}, {Dupac}, {Dusini}, {Efstathiou}, {Elsner}, {En{\ss}lin}, {Eriksen}, {Fantaye}, {Farhang}, {Fergusson}, {Fernandez-Cobos}, {Finelli}, {Forastieri}, {Frailis}, {Fraisse}, {Franceschi}, {Frolov}, {Galeotta}, {Galli}, {Ganga}, {G{\'e}nova-Santos}, {Gerbino}, {Ghosh}, {Gonz{\'a}lez-Nuevo}, {G{\'o}rski}, {Gratton}, {Gruppuso}, {Gudmundsson}, {Hamann}, {Handley}, {Hansen}, {Herranz}, {Hildebrandt}, {Hivon}, {Huang}, {Jaffe}, {Jones}, {Karakci}, {Keih{\"a}nen},
  {Keskitalo}, {Kiiveri}, {Kim}, {Kisner}, {Knox}, {Krachmalnicoff}, {Kunz}, {Kurki-Suonio}, {Lagache}, {Lamarre}, {Lasenby}, {Lattanzi}, {Lawrence}, {Le Jeune}, {Lemos}, {Lesgourgues}, {Levrier}, {Lewis}, {Liguori}, {Lilje}, {Lilley}, {Lindholm}, {L{\'o}pez-Caniego}, {Lubin}, {Ma}, {Mac{\'\i}as-P{\'e}rez}, {Maggio}, {Maino}, {Mandolesi}, {Mangilli}, {Marcos-Caballero}, {Maris}, {Martin}, {Martinelli}, {Mart{\'\i}nez-Gonz{\'a}lez}, {Matarrese}, {Mauri}, {McEwen}, {Meinhold}, {Melchiorri}, {Mennella}, {Migliaccio}, {Millea}, {Mitra}, {Miville-Desch{\^e}nes}, {Molinari}, {Montier}, {Morgante}, {Moss}, {Natoli}, {N{\o}rgaard-Nielsen}, {Pagano}, {Paoletti}, {Partridge}, {Patanchon}, {Peiris}, {Perrotta}, {Pettorino}, {Piacentini}, {Polastri}, {Polenta}, {Puget}, {Rachen}, {Reinecke}, {Remazeilles}, {Renzi}, {Rocha}, {Rosset}, {Roudier}, {Rubi{\~n}o-Mart{\'\i}n}, {Ruiz-Granados}, {Salvati}, {Sandri}, {Savelainen}, {Scott}, {Shellard}, {Sirignano}, {Sirri}, {Spencer}, {Sunyaev}, {Suur-Uski}, {Tauber}, {Tavagnacco},
  {Tenti}, {Toffolatti}, {Tomasi}, {Trombetti}, {Valenziano}, {Valiviita}, {Van Tent}, {Vibert}, {Vielva}, {Villa}, {Vittorio}, {Wandelt}, {Wehus}, {White}, {White}, {Zacchei}, \& {Zonca}}]{Planck2020}
{Planck Collaboration}, {Aghanim}, N., {Akrami}, Y., {et~al.} 2020, \aap, 641, A6

\bibitem[{{Platts} {et~al.}(2019){Platts}, {Weltman}, {Walters}, {Tendulkar}, {Gordin}, \& {Kandhai}}]{platts2018}
{Platts}, E., {Weltman}, A., {Walters}, A., {et~al.} 2019, \physrep, 821, 1

\bibitem[{{Popov} \& {Postnov}(2010)}]{Popov}
{Popov}, S.~B. \& {Postnov}, K.~A. 2010, in Evolution of Cosmic Objects through their Physical Activity, ed. H.~A. {Harutyunian}, A.~M. {Mickaelian}, \& Y.~{Terzian}, 129--132

\bibitem[{{Ransom}(2011)}]{PRESTO}
{Ransom}, S. 2011, {PRESTO: PulsaR Exploration and Search TOolkit}, Astrophysics Source Code Library, record ascl:1107.017

\bibitem[{{Ridnaia} {et~al.}(2021){Ridnaia}, {Svinkin}, {Frederiks}, {Bykov}, {Popov}, {Aptekar}, {Golenetskii}, {Lysenko}, {Tsvetkova}, {Ulanov}, \& {Cline}}]{Ridnaia}
{Ridnaia}, A., {Svinkin}, D., {Frederiks}, D., {et~al.} 2021, Nature Astronomy, 5, 372

\bibitem[{{Scholz} {et~al.}(2017){Scholz}, {Bogdanov}, {Hessels}, {Lynch}, {Spitler}, {Bassa}, {Bower}, {Burke-Spolaor}, {Butler}, {Chatterjee}, {Cordes}, {Gourdji}, {Kaspi}, {Law}, {Marcote}, {McLaughlin}, {Michilli}, {Paragi}, {Ransom}, {Seymour}, {Tendulkar}, \& {Wharton}}]{Scholz2017}
{Scholz}, P., {Bogdanov}, S., {Hessels}, J.~W.~T., {et~al.} 2017, \apj, 846, 80

\bibitem[{{Scholz} {et~al.}(2020){Scholz}, {Cook}, {Cruces}, {Hessels}, {Kaspi}, {Majid}, {Naidu}, {Pearlman}, {Spitler}, {Bandura}, {Bhardwaj}, {Cassanelli}, {Chawla}, {Gaensler}, {Good}, {Josephy}, {Karuppusamy}, {Keimpema}, {Kirichenko}, {Kirsten}, {Kocz}, {Leung}, {Marcote}, {Masui}, {Mena-Parra}, {Merryfield}, {Michilli}, {Naudet}, {Nimmo}, {Pleunis}, {Prince}, {Rafiei-Ravandi}, {Rahman}, {Shin}, {Smith}, {Stairs}, {Tendulkar}, \& {Vanderlinde}}]{Scholz2020}
{Scholz}, P., {Cook}, A., {Cruces}, M., {et~al.} 2020, \apj, 901, 165

\bibitem[{{Scholz} {et~al.}(2016){Scholz}, {Spitler}, {Hessels}, {Chatterjee}, {Cordes}, {Kaspi}, {Wharton}, {Bassa}, {Bogdanov}, {Camilo}, {Crawford}, {Deneva}, {van Leeuwen}, {Lynch}, {Madsen}, {McLaughlin}, {Mickaliger}, {Parent}, {Patel}, {Ransom}, {Seymour}, {Stairs}, {Stappers}, \& {Tendulkar}}]{Scholz2016}
{Scholz}, P., {Spitler}, L.~G., {Hessels}, J.~W.~T., {et~al.} 2016, \apj, 833, 177

\bibitem[{{Seymour} {et~al.}(2019){Seymour}, {Michilli}, \& {Pleunis}}]{dmphase}
{Seymour}, A., {Michilli}, D., \& {Pleunis}, Z. 2019, {DM\_phase: Algorithm for correcting dispersion of radio signals}, Astrophysics Source Code Library, record ascl:1910.004

\bibitem[{{Shin} \& {CHIME/FRB Collaboration}(2024)}]{AtelChimeFRB20240114A}
{Shin}, K. \& {CHIME/FRB Collaboration}. 2024, The Astronomer's Telegram, 16420

\bibitem[{{Snelders} {et~al.}(2024){Snelders}, {Bhandari}, {Kirsten}, {Hessels}, {Marcote}, {Hewitt}, {Gawronski}, {Puchalska}, {Ould-Boukattine}, {Gopinath}, {Nimmo}, {Karuppusamy}, {Herrmann}, {Yang}, {Blaauw}, {Buttaccio}, {Maccaferri}, {Bach}, {Feiler}, {Bray}, {Williams}, {Wrigley}, {Keimpema}, {Paragi}, {Burgay}, {Corongiu}, {Giroletti}, {Kramer}, {Pilia}, {Spitler}, {Surcis}, {Trudu}, {Yuan}, {Wang}, \& {Bezrukovs}}]{EVNFRB20240114A}
{Snelders}, M.~P., {Bhandari}, S., {Kirsten}, F., {et~al.} 2024, The Astronomer's Telegram, 16542

\bibitem[{Spitler {et~al.}(2012)Spitler, Cordes, Chatterjee, \& Stone}]{Spitler_2012}
Spitler, L.~G., Cordes, J.~M., Chatterjee, S., \& Stone, J. 2012, \apj, 748, 73

\bibitem[{{Tavani} {et~al.}(2021){Tavani}, {Casentini}, {Ursi}, {Verrecchia}, {Addis}, {Antonelli}, {Argan}, {Barbiellini}, {Baroncelli}, {Bernardi}, {Bianchi}, {Bulgarelli}, {Caraveo}, {Cardillo}, {Cattaneo}, {Chen}, {Costa}, {Del Monte}, {Di Cocco}, {Di Persio}, {Donnarumma}, {Evangelista}, {Feroci}, {Ferrari}, {Fioretti}, {Fuschino}, {Galli}, {Gianotti}, {Giuliani}, {Labanti}, {Lazzarotto}, {Lipari}, {Longo}, {Lucarelli}, {Magro}, {Marisaldi}, {Mereghetti}, {Morelli}, {Morselli}, {Naldi}, {Pacciani}, {Parmiggiani}, {Paoletti}, {Pellizzoni}, {Perri}, {Perotti}, {Piano}, {Picozza}, {Pilia}, {Pittori}, {Puccetti}, {Pupillo}, {Rapisarda}, {Rappoldi}, {Rubini}, {Setti}, {Soffitta}, {Trifoglio}, {Trois}, {Vercellone}, {Vittorini}, {Giommi}, \& {D'Amico}}]{Tavani2021}
{Tavani}, M., {Casentini}, C., {Ursi}, A., {et~al.} 2021, Nature Astronomy, 5, 401

\bibitem[{{Tian} {et~al.}(2024){Tian}, {Rajwade}, {Pastor-Marazuela}, {Stappers}, {Bezuidenhout}, {Caleb}, {Jankowski}, {Barr}, \& {Kramer}}]{MeerKATLocalization}
{Tian}, J., {Rajwade}, K.~M., {Pastor-Marazuela}, I., {et~al.} 2024, \mnras, 533, 3174

\bibitem[{{Trudu} {et~al.}(2023){Trudu}, {Pilia}, {Nicastro}, {Guidorzi}, {Orlandini}, {Zampieri}, {Marthi}, {Ambrosino}, {Possenti}, {Burgay}, {Casentini}, {Mereminskiy}, {Savchenko}, {Palazzi}, {Panessa}, {Ridolfi}, {Verrecchia}, {Anedda}, {Bernardi}, {Bachetti}, {Burenin}, {Burtovoi}, {Casella}, {Fiori}, {Frontera}, {Gajjar}, {Gardini}, {Ge}, {Guijarro-Rom{\'a}n}, {Ghedina}, {Hermelo}, {Jia}, {Li}, {Liao}, {Li}, {Lu}, {Lutovinov}, {Naletto}, {Ochner}, {Papitto}, {Perri}, {Pittori}, {Safonov}, {Semena}, {Strakhov}, {Tavani}, {Ursi}, {Xiong}, {Zhang}, \& {Zheltoukhov}}]{Trudu2023}
{Trudu}, M., {Pilia}, M., {Nicastro}, L., {et~al.} 2023, \aap, 676, A17

\bibitem[{{Tub{\'\i}n-Arenas} {et~al.}(2024){Tub{\'\i}n-Arenas}, {Krumpe}, {Lamer}, {Haase}, {Sanders}, {Brunner}, {Homan}, {Schwope}, {Georgakakis}, {Poppenhaeger}, {Traulsen}, {K{\"o}nig}, {Merloni}, {Gueguen}, {Strong}, \& {Liu}}]{Tubin}
{Tub{\'\i}n-Arenas}, D., {Krumpe}, M., {Lamer}, G., {et~al.} 2024, \aap, 682, A35

\bibitem[{{Uttarkar} {et~al.}(2024){Uttarkar}, {Kumar}, {Lower}, \& {Shannon}}]{ATel16430}
{Uttarkar}, P.~A., {Kumar}, P., {Lower}, M.~E., \& {Shannon}, R.~M. 2024, The Astronomer's Telegram, 16430

\bibitem[{{van Straten} \& {Bailes}(2011)}]{dspsr}
{van Straten}, W. \& {Bailes}, M. 2011, \pasa, 28, 1

\bibitem[{{van Straten} {et~al.}(2012){van Straten}, {Demorest}, \& {Oslowski}}]{psrchive}
{van Straten}, W., {Demorest}, P., \& {Oslowski}, S. 2012, Astronomical Research and Technology, 9, 237

\bibitem[{{Williams} {et~al.}(2017){Williams}, {Clavel}, {Newton}, \& {Ryzhkov}}]{pwkit}
{Williams}, P. K.~G., {Clavel}, M., {Newton}, E., \& {Ryzhkov}, D. 2017, {pwkit: Astronomical utilities in Python}, Astrophysics Source Code Library, record ascl:1704.001

\bibitem[{{Yan} {et~al.}(2024){Yan}, {Yu}, {Page}, {Lin}, {Li}, {Niu}, {Law}, {Zhang}, {Chatterjee}, {Zhang}, \& {Anna-Thomas}}]{Yan2024}
{Yan}, Z., {Yu}, W., {Page}, K.~L., {et~al.} 2024, arXiv e-prints, arXiv:2402.12084

\bibitem[{{Zhang}(2018)}]{Zhang2018}
{Zhang}, B. 2018, \apjl, 867, L21

\bibitem[{{Zhang} {et~al.}(2024{\natexlab{a}}){Zhang}, {Wu}, {Cao}, {Zhu}, {Zhang}, {Niu}, {Xie}, {Zhou}, {Wang}, {Zhu}, {Zhang}, {Wang}, {Niu}, {Di Li}, {Han}, {Lee}, {Wang}, {Gao}, {Feng}, {Jiang}, {Jing}, {Li}, {Lu}, {Luo}, {Lyu}, {Wang}, {Xu}, {Yang}, {Yu}, {Zhang}, \& {Project}}]{ATel16505}
{Zhang}, J., {Wu}, Q., {Cao}, S., {et~al.} 2024{\natexlab{a}}, The Astronomer's Telegram, 16505

\bibitem[{{Zhang} {et~al.}(2024{\natexlab{b}}){Zhang}, {Zhu}, {Cao}, {Zhou}, {Zhang}, {Xie}, {Wu}, {Wang}, {Wang}, {Niu}, {Di Li}, {Zhu}, {Zhang}, {Han}, {Lee}, {Wang}, {Gao}, {Feng}, {Jiang}, {Jing}, {Li}, {Lu}, {Luo}, {Lyu}, {Wang}, {Xu}, {Yang}, {Yu}, {Zhang}, \& {Project}}]{ATel16433}
{Zhang}, J., {Zhu}, Y., {Cao}, S., {et~al.} 2024{\natexlab{b}}, The Astronomer's Telegram, 16433

\bibitem[{{Zhong} {et~al.}(2024){Zhong}, {Li}, {Zhang}, \& {Dai}}]{Zhong2024}
{Zhong}, S.-Q., {Li}, L., {Zhang}, B., \& {Dai}, Z.-G. 2024, \apj, 976, 52

\bibitem[{{Zhu} {et~al.}(2023){Zhu}, {Xu}, {Zhou}, {Lin}, {Wang}, {Wang}, {Zhang}, {Niu}, {Chen}, {Li}, {Meng}, {Lee}, {Zhang}, {Feng}, {Ge}, {G{\"o}{\u{g}}{\"u}{\c{s}}}, {Guan}, {Han}, {Jiang}, {Jiang}, {Kouveliotou}, {Li}, {Miao}, {Miao}, {Men}, {Niu}, {Wang}, {Wang}, {Xu}, {Xu}, {Xue}, {Yang}, {Yu}, {Yuan}, {Yue}, {Zhang}, \& {Zhang}}]{zhu}
{Zhu}, W., {Xu}, H., {Zhou}, D., {et~al.} 2023, Science Advances, 9, eadf6198

\end{thebibliography}

\begin{appendix}%First appendix

\section{Effelsberg UBB data analysis}
\label{appendix:ubb}
 The UBB receiver delivers \texttt{PSRFITS} files for every sub-band, which were  searched for single bursts using \texttt{TransientX} \citep{TransientX} separately in every sub-band. We performed a search for short bursts with a maximum width of 10\,ms on the full resolution data from the UBB. Additionally, we ran a search for longer bursts with a maximum width of 100\,ms with a time down sampling factor of 10. The detected candidates were filtered using the \texttt{replot\_fil} command within \texttt{TransientX} using an S/N cutoff of 7 and then inspected manually to identify false positive detections. \review{After discarding false positives, we matched the candidates detected by the long and short searches within each sub-band, i.e. if the same burst was detected in both searches, we kept only the information about the detection with higher S/N. This results in 186 detected bursts in band 1 and band 2 each, 133 bursts in band 3, 32 bursts in band 4, and 10 bursts in band 5.} We corrected the arrival times from the \review{individual sub-}band \review{detections} to infinite frequency and matched any overlapping candidates to avoid \review{counting} the same burst \review{multiple times in case it was} detected in \review{several sub-}bands. \review{This leads to a total number of 459 individual bursts across all sub-bands, whose arrival times are evenly distributed across the observation}. We extracted 1\,s pulsar archive snippets around the burst arrival times using \texttt{DSPSR} \citep{dspsr}. Flux and polarization calibration was performed using the \texttt{pac} and \texttt{fluxcal} programs within \texttt{PSRCHIVE} \citep{psrchive} for every sub-band separately. Calibration solutions were derived from observations of 3C\,48 on 9 May 2024 using the \cite{perleybutler2017} flux scale. To be conservative, we assume a 5\,\% error on the flux scale, which is on the high-end of the suggested error by \cite{perleybutler2017}. Prior to flux calibration, automatic radio frequency interference (RFI) cleaning was performed on the archive files using \texttt{clfd} \citep{clfd}, as well as additional manual flagging using the \texttt{PSRCHIVE} tool \texttt{pazi}. To combine the different sub-bands files from the UBB we loaded the calibrated and RFI-flagged archive files into the \texttt{PSRCHIVE} python framework and combined them as \texttt{numpy} arrays. Our code to combine the different sub-bands of the UBB is available on GitHub\footnote{\url{https://github.com/flep198/ubb_tools}}.

 For further analysis, we dedispersed all bursts using a common DM value of $(527.979\pm 0.085)\,\textrm{pc}\,\textrm{cm}^{-3}$. This DM value was derived from burst B1, which shows the most complex temporal substructure among the three brightest bursts, using the \texttt{DM\_PHASE} package \citep{dmphase}. We assume that the DM of FRB\,20240114A does not vary significantly within the observation duration, which is valid for other repeating FRBs \citep[e.g.,][]{FRB121102Hessels}. In a follow-up study, we will address this question for FRB\,20240114A in more detail and study possible DM variations between the 459 detected bursts.
The topocentric arrival times of the bursts were corrected to infinite frequency and to the arrival time at the solar system barycenter using \texttt{get\_barycentric\_toas} within \texttt{PINT} \citep{PINT} and the JPL DE421 ephemeris. We note that the uncertainty of the DM and the exact position of the FRB can introduce an error on the order of several milliseconds.

\section{Astrometric correction for XMM-Newton data}
\label{appendix:eposcorr}

To determine the exact position of FRB\,20240114A in the \textsl{XMM-Newton} images, we performed a source detection in the full PN and MOS data of every observation using the \texttt{edetect\_stack} task in \texttt{SAS}. While there is no significant source detected in the small-window region of the chips, the outer regions of the MOS-chip still allow for the detection of several field sources. The detected X-ray sources were matched with the Gaia DR3 catalog \citep{GaiaDR3} using \texttt{eposcorr}, as described by \cite{kurpas2024}. The resulting astrometric corrections in Right Ascension $\Delta$RA and in declination $\Delta$Dec are shown in Table\,\ref{tab:eposcorr} and were applied to each dataset. %To our knowledge, for previous \textsl{XMM-Newton} studies of FRBs \citep[e.g.,][]{Scholz2017,Pilia2020,Pearlman2023,Cook2024} this astrometric correction was not carried out. 
The corrections are small compared to the extraction radius. In order to test if they lead to a significant change in our results, we performed all of our analysis also without applying this correction, which indeed has no significant effect on the results since the corrections are small compared to the PSF.

\begin{table}
\caption{Astrometric corrections for the \textsl{XMM-Newton} pointings calculated with \texttt{eposcorr}.}
\label{tab:eposcorr}
\centering
\begin{tabular}{ccc}
\hline\hline
{Obs.ID}  & {$\Delta$RA} & {$\Delta$Dec}  \\
  {}   &  {["]}  & {["]}  \\
\hline
0935190601 & $1.08\pm 0.11$ & $0.58\pm 0.17$  \\
0935190701 & $-0.115\pm 0.021$ & $0.555\pm 0.060$ \\
0935191001 & $1.6195\pm 0.0039$ & $0.740\pm 0.025$ \\
\hline
\end{tabular}
\end{table}
\end{appendix}

% WARNING
%-------------------------------------------------------------------
% Please note that we have included the references to the file aa.dem in
% order to compile it, but we ask you to:
%
% - use BibTeX with the regular commands:
%   
%
% - join the .bib files when you upload your source files
%-------------------------------------------------------------------

\end{document}